\documentclass[preprint]{aastex}

\newcommand\ha{H$\alpha$}
\newcommand\ca{\mbox{Ca\,{\sc ii}}}
\newcommand\mg{\mbox{Mg\,{\sc ii}}}
\newcommand\cb{\mbox{C\,{\sc ii}}}
\newcommand\si{\mbox{Si\,{\sc iv}}}
\newcommand{\kms}{~km~s$^{-1}$}

\begin{document}

\title{Oscillatory Response of the Solar Chromosphere to a Strong Downflow Event above a Sunspot}

\author{Hannah Kwak, Jongchul Chae, Donguk Song}
\affil{Astronomy Program, Department of Physics and Astronomy,
 Seoul National University, Seoul 151-742, Korea}

\and

\author{Yeon-Han Kim, Eun-Kyung Lim}
\affil{Korea Astronomy and Space Science Institute, Daejeon 305-348, Korea}

\and

\author{Maria S. Madjarska}
\affil{Armagh Observatory, College Hill, Armagh BT61 9DG, N. Ireland}

\begin{abstract}

We report three-minute oscillations in the solar chromosphere driven by a strong downflow event in a sunspot. We used the Fast Imaging Solar Spectrograph of the 1.6~m New Solar Telescope and the Interface Region Imaging Spectrograph (IRIS). The strong downflow event is identified in the chromospheric and transition region lines above the sunspot umbra. After the event, oscillations occur at the same region. The amplitude of the Doppler velocity oscillations is 2~\kms, and gradually decreases with time. In addition, the period of the oscillations gradually increases from 2.7~minutes to 3.3~minutes. In the IRIS 1330 slit-jaw images, we identify a transient brightening near the footpoint of the downflow detected in the \ha+0.5~\AA\ image. The characteristics of the downflowing material are consistent with those of sunspot plumes. Based on our findings, we suggest that the gravitationally stratified atmosphere came to oscillate with three minute period in response to the impulsive downflow event as was theoretically investigated by \citet{chae15}.

\end{abstract}


\section{INTRODUCTION}

Ever since discovery of three-minute umbral oscillations \citep{bec69, bec72,gio72}, they have been studied for several decades by numerous authors \citep[for reviews, see][]{sta99, bog06, kho15}. The oscillations are detected in the multiple layers of the solar atmosphere above sunspot umbrae \citep[e.g.,][]{lites85, tho87, mal99, osh02}. They are generally manifested as periodic fluctuations of intensity and velocity. The oscillations are commonly regarded as slow magnetoacoustic waves upwardly propagating in the gravitationally stratified medium, as the oscillations in the higher atmosphere lag behind those in the lower atmosphere \citep{bry99, bry04, tian14}. It is believed that the power of the three-minute oscillations may mostly come from the photosphere or below it. Two specific processes are currently considered as the major candidates for the driver of the oscillations: $p$-mode absorption of global solar oscillations and magnetoconvection inside sunspots \citep[see review by][and references therein]{kho15}.

Theory predicts that the three-minute umbral oscillations can be excited by impulsive disturbances inside the atmosphere as well. Early studies by \cite{lamb09} and \cite{kal94} showed that the atmosphere disturbed by an impulsive disturbance at a point begins to oscillate with frequency asymptotically approaching the acoustic cutoff frequency. The recent study of \cite{chae15} found that the necessary condition for the cutoff frequency oscillations to have enough power is the occurrence of the impulsive disturbance of large vertical extent. Since there are many activities in the chromosphere, we expect that impulsive disturbances that produce the three-minute oscillations occur even though they may not be the major driver of such oscillations. Despite this expectation, so far there has been no observational report of such three-minute oscillations generated by impulsive events in the chromosphere above sunspots. This is in contrast with the common findings of oscillations and waves in the corona driven by the strongest impulsive events such as flares \citep{asc99, nak99, sch02}, lower coronal eruptions/ejections \citep{zim15}, and filament destabilizations \citep{sch02}. Strong X-class flares even affect the solar interior and cause sunquakes which are acoustic waves produced in the photosphere \citep{kos98, kos06}. Probably the impulsive events leading to three-minute oscillations in the chromosphere may be too small and too weak to be detected.

In this letter, for the first time we report on the occurrence of the three-minute oscillations driven by an impulsive disturbance in the chromosphere. In this specific study, the disturbance was caused by a strong downflow event detected in the chromospheric and transition region (TR) lines above a sunspot umbra. It is important to understand the properties of the oscillations and waves in order to estimate physical quantities of the medium for further study. We analyze multi-wavelength data acquired by the Fast Imaging Solar Spectrograph \citep[FISS;][]{chae13} installed at the 1.6~m New Solar Telescope (NST) of Big Bear Solar Observatory and Interface Region Imaging Spectrograph \citep[IRIS;][]{dep14}.

\section{OBSERVATIONS AND DATA ANALYSIS}

We observed a small sunspot in NOAA active region 12172 on 2014 September 27. It was located between a leading main sunspot and a trailing main sunspot. The region of our interest was so variable that the morphology of the small sunspot kept changing within several hours, and the sunspot finally disappeared the day after our observation period. The FISS is a dual-band Ech\'{e}lle spectrograph that records the \ha\ band and the \ca\ 8542~\AA\ band simultaneously with imaging capability. Imaging is done by fast scanning of a slit over the field of view (FOV) and the step size is 0.\arcsec16. With this instrument we acquired data for an hour from 17:03:40 to 18:00:50~UT with the aid of the adaptive optics system. The spatial sampling along the slit is 0.\arcsec16 and the spectral samplings are 0.019~\AA\ and 0.026~\AA\ at the \ha\ line and the \ca\ line, respectively. The FOV of the raster image is 24\arcsec $\times$ 40\arcsec\ and the time cadence is 22~s.

The IRIS observation was done with large coarse 2-step raster mode from 11:29:36 to 17:36:50~UT and we analyzed the data taken from 17:03:45 to 17:36:29~UT which correspond to the overlapping period with the FISS data. The IRIS spectra were acquired in the near-ultraviolet (NUV, 2783--2834~\AA) and far-ultraviolet (FUV, 1332--1358~\AA, 1390--1406~\AA) band, and we utilized \mg\ k 2796~\AA\ (formed in the chromosphere), \cb\ 1336~\AA\ (formed in the lower TR) and \si\ 1403~\AA\ (formed in the middle TR) lines for our study. In addition, the IRIS slit-jaw images (SJIs) were taken with the filters of SJI 2796~\AA\ and SJI 1330~\AA. The spatial pixel size is 0.\arcsec33 and the time cadence is 18~s at both of the spectral data and SJIs. We used the IRIS level 2 data which are dark current subtracted, flat-fielded, geometric corrected and wavelength calibrated. Additionally, we used data from the Atmospheric Imaging Assembly \citep[AIA;][]{lemen12} on board the \textit{Solar Dynamics Observatory} (\textit{SDO}) to investigate higher atmospheric responses.

\begin{figure}
\plotone{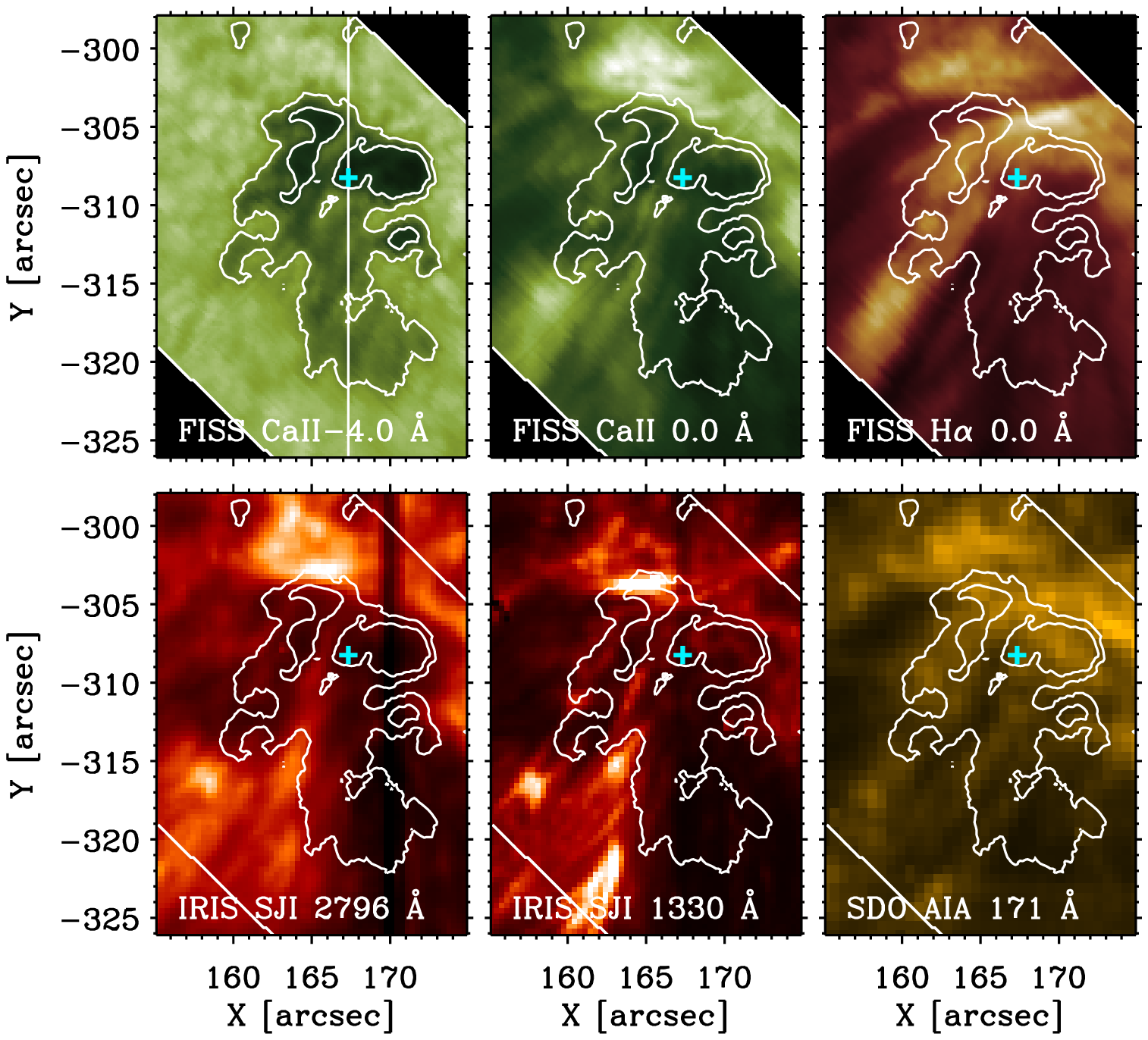}
\caption{Images of the small sunspot taken by the FISS, IRIS and the SDO at 17:30~UT on 2014 September 27. The white vertical line in the FISS~\ca-4~\AA\ image indicates the location of the IRIS slit position, and the cyan cross symbols mark the selected position for detailed analysis in Figure~\ref{wt} and Figure~\ref{losv}. \label{fig1}}
\end{figure}

Figure~\ref{fig1} shows the small sunspot at different solar atmospheric layers from the upper photosphere to the corona. The FISS \ca\ and \ha\ raster images show the low atmosphere from the upper photosphere (line wings) to the chromosphere (line core). The IRIS~SJI~2796~\AA\ and SJI 1330~\AA\ correspond to the chromosphere and the lower TR, respectively, and the AIA 171~\AA\ corresponds to the upper TR and the corona. Alignment of the FISS and the IRIS data was carried out by using the FISS \ca\ 8542~\AA\ raster image (Figure~\ref{fig1} top center panel) and the IRIS~SJI~2796~\AA\ image (Figure~\ref{fig1} bottom left panel).

We inferred the line-of-sight (LOS) Doppler velocity by applying single Gaussian fit and the lambdameter method to the line profiles. Generally, \mg\ and \cb\ line profiles are central-reversed (double-peaked), but inside the sunspot, they are single-peaked and fitted well by the single Gaussian fits \citep{tian14}. Similarly, since the \ca\ absorption line profiles in sunspot umbrae have an emission core, the core of the line profile is appropriate for applying the single Gaussian fit. Thus, we applied the single Gaussian fits to the line profiles of the \ca, \mg, \cb\ and \si\ lines. Whereas, in the case of the \ha\ absorption line profile, we applied the lambdameter method, which calculates the Doppler velocity by determining the mid-point of $2\Delta$$\lambda$ \citep{deub96}. We set the $\Delta$$\lambda$ as 0.5~\AA\ in this study.

Since the \ha\ line has a broad absorption core, spectral and temporal variations are not conspicuous in the \ha\ line profile. Therefore, in order to see the variations clearly, we defined a contrast profile as $C_{\lambda} = (I_{\lambda}-I_{ref})/I_{ref}$, where $I_{\lambda}$ is the spectral intensity profile at the selected position, and $I_{ref}$ is the reference profile which is a temporally-averaged intensity profile at the same position during our observation period. Using the contrast profiles of the \ha\ and \ca~8542~\AA\ lines, wavelength-time ($\lambda$$-$$t$) plots of the intensity contrast are made in Figure~\ref{wt}~(f)-(g).

\section{RESULTS}

\begin{figure}
\plotone{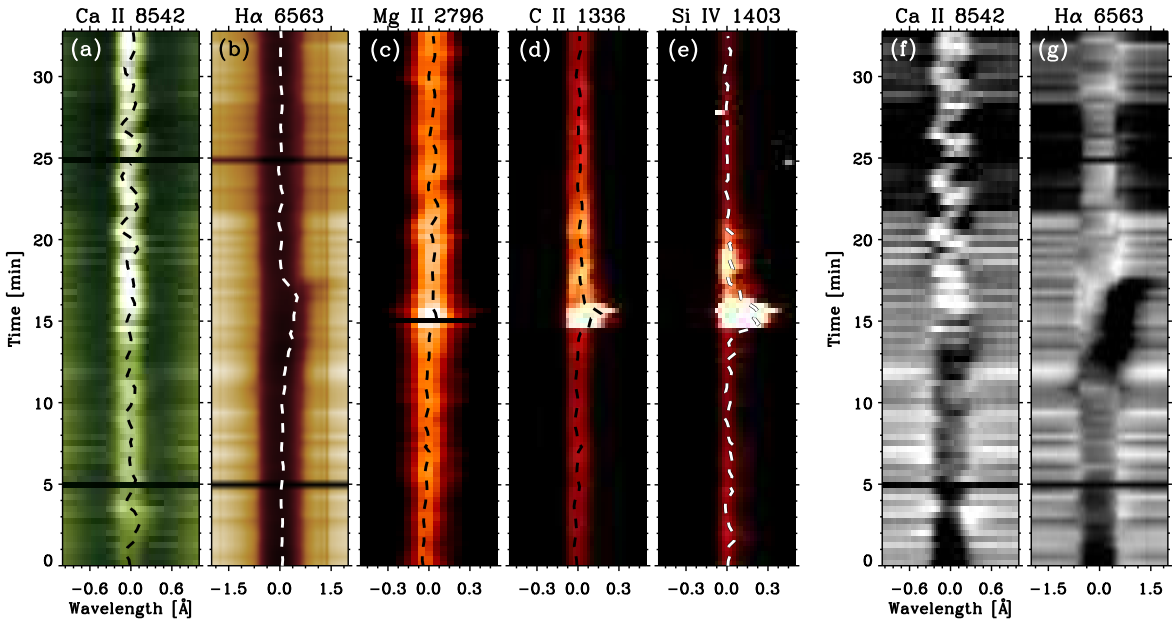}
\caption{Panels (a)-(e) : Wavelength-time ($\lambda$$-$$t$) plots of the five spectral lines at the position marked with cyan cross symbols in Figure~\ref{fig1}. The dashed lines represent the track of the line center obtained by applying the single Gaussian fits to each line (magnified by a factor of 3). Panels (f)-(g) : $\lambda$$-$$t$ plots of the \ca\ and \ha\ intensity contrast. Time $t=0$ corresponds to 17:03:40~UT. \label{wt}}
\end{figure}

The remarkable finding of our study is the detection of a strong downflow event and the associated oscillations on the same region. Figure~\ref{wt} shows the $\lambda$$-$$t$ plots at a selected position which is marked with a cyan cross symbol in Figure~\ref{fig1}. The $\lambda$$-$$t$ plots represent the spectral and temporal variations in the \ca, \ha, \mg, \cb\ and \si\ lines. In the plots of the \cb\ and \si\ lines, a striking red-shifted feature suddenly appeared at $t$ $\approx$ $14$ mins. The red-shifted feature is also identified in the $\lambda$$-$$t$ plots of the \ha\ line, and it stands out more clearly in the $\lambda$$-$$t$ plot of the \ha\ intensity contrast (see Figure~\ref{wt}~(g)). We can see that the red-shifted feature showed up first in the \ha\ line and then in the other lines. In the $\lambda$$-$$t$ plot of the \ha\ intensity contrast, the feature began to appear at $t$ $\approx$ $10$ mins, and the red shift gradually increased. Four minutes later, in the \cb\ and \si\ lines, the red-shifted feature abruptly showed up and the red shift rapidly increased. This red-shifted feature eventually represents a downflow event occurring in the solar chromosphere and TR.

Interestingly, after the abrupt appearance of the red-shifted feature, oscillation patterns were identified in all of the $\lambda$$-$$t$ plots. The red-shifted feature disappeared at $t$ $\approx$ $17$ mins in the $\lambda$$-$$t$ plot of the \ha\ intensity contrast (see Figure~\ref{wt}~(g)), and the oscillations began immediately at around $t=18$ mins persisting until the end of our observation period or even more. Note that weak oscillations are present even before the downflow event, but they are strongly enhanced after the downflow event. The oscillations are conspicuous in the \ca\ and \mg\ lines, and they are even clearer in the \ca\ $\lambda$$-$$t$ plot of intensity contrast (see Figure~\ref{wt}~(f)). Additionally, the intensity of each line slightly increased after the downflow event, as can be seen in the $\lambda$$-$$t$ plots of the \ca, \mg, \cb\ and \si\ lines, and more evident in the $\lambda$$-$$t$ plot of the \ca\ and \ha\ intensity contrast. It suggests that the downflow event caused the oscillations and the associated heating in this region.

\begin{figure}
\plotone{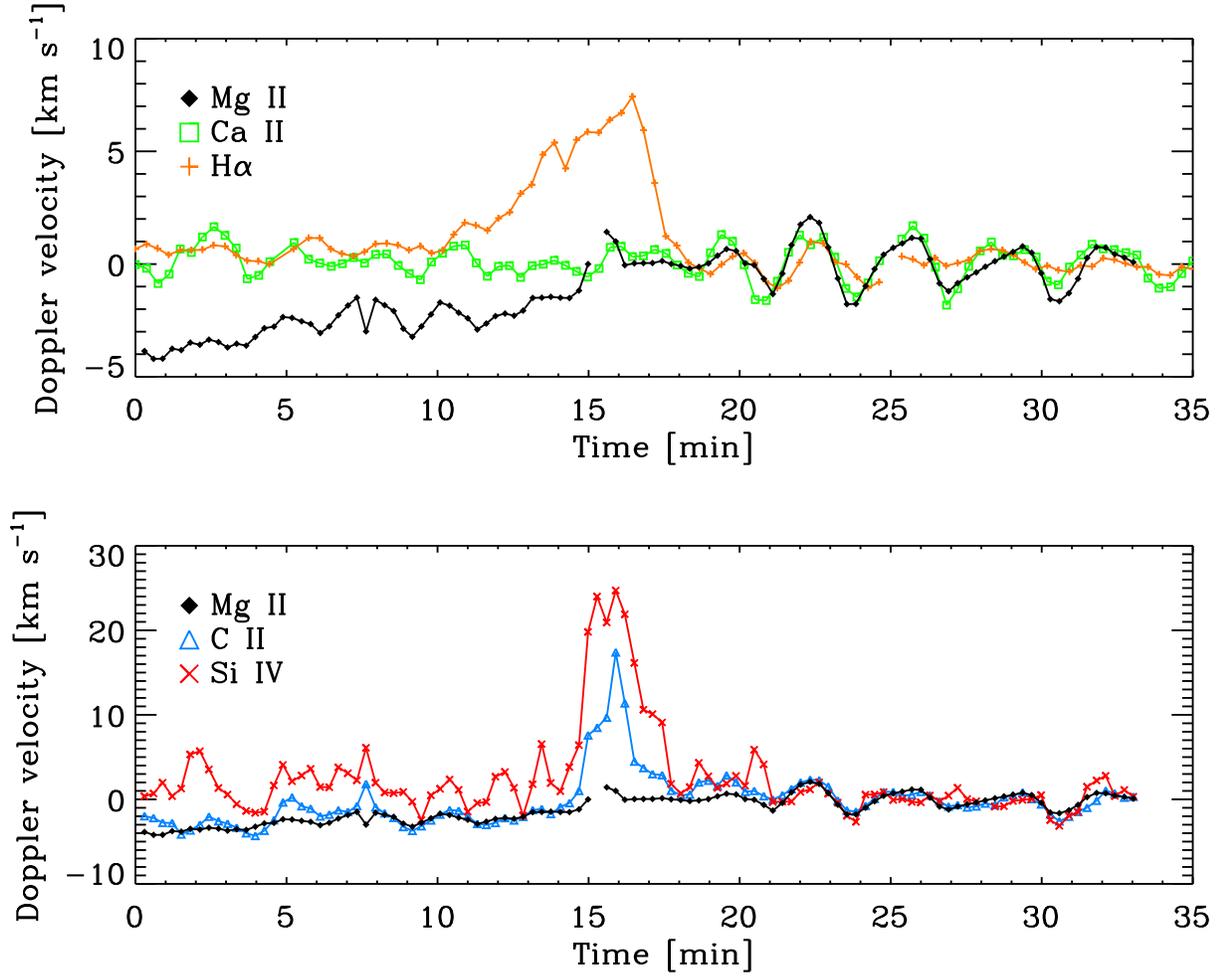}
\caption{Temporal variations of the LOS Doppler velocity in the \ca, \ha\ and \mg\ lines (upper panel), and in the \mg, \cb\ and \si\ lines (lower panel). Time $t=0$ corresponds to 17:03:40~UT. \label{losv}}
\end{figure}

Figure~\ref{losv} shows the temporal variations of the LOS Doppler velocity at the same position. The downflow is identified in the \ha, \cb\ and \si\ lines in common, but the temporal variations of the Doppler velocity are different in each of the lines. In the \ha\ line, the downflow event began at $t=10$ mins, and the Doppler velocity of the downflow gradually increased reaching its peak value of 7.8~\kms\ at $t=16.5$~mins. Meanwhile, in the \cb\ and \si\ lines, the event began at $t=14$~mins and the Doppler velocity sharply increased reaching their peaks at $t=16$~mins with values of 18.5~\kms\ and 27.2~\kms, respectively. We expect the true Doppler velocity of the downflow may be a little higher than these values since the single Gaussian fits may include stationary component as well.

We find several properties of the oscillations that occurred after the downflow event from Figure~\ref{losv}. The oscillations are seen in all of the lines, and it implies that the oscillations are detected in a wide range of temperature, i.e., from the chromosphere temperature ($10^{4}$~K) to the middle TR temperature ($10^{4.9}$~K). The amplitude of the oscillations is about 2~\kms\ in most of the lines and slightly decreases with time. From the wavelet analysis of power spectrum \citep{tor98}, we find that the dominant period of the oscillations is initially 2.7~minutes and gradually increases to 3.3~minutes. These properties of the oscillations are well identified in the \mg\ line, and other lines as well. In the case of the \ca\ and \cb\ lines, the oscillation patterns are very similar to those of the \mg\ line. The \si\ line also shows the same oscillations, but the oscillation patterns are less conspicuous in the \si\ line. The amplitude of the \ha\ Doppler velocity appears smaller than that of the other four lines which may be because the lambdameter method underestimates Doppler velocity of features in the upper chromosphere. Nevertheless, we can still identify the oscillation patterns in the \ha\ line.

Another interesting phenomenon is the coherency of the LOS Doppler velocity patterns in the five different spectral lines after the downflow event (see Figure~\ref{losv}). Before the event, each line shows its common velocity reflecting Doppler velocity at each line formation temperature. For example, in the \mg\ line, upward motion is dominant and in the \si\ line downward motion is dominant. After the event, however, every line shows a similar pattern of the Doppler velocity. The amplitudes of each line are almost the same and there are no clear phase differences among the five spectral lines.

\begin{figure}
\plotone{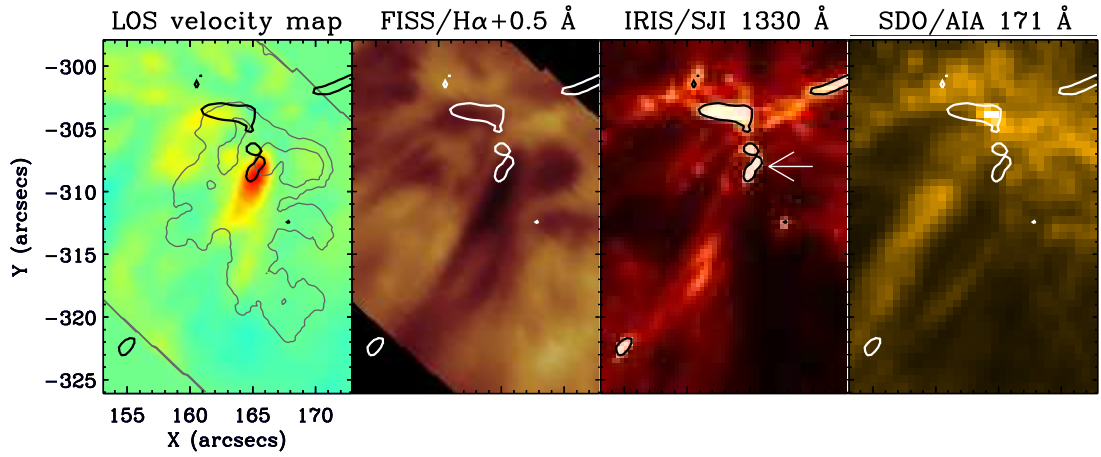}
\caption{Images of the FISS \ha\ LOS Doppler velocity map, \ha+0.5~\AA, IRIS~SJI~1330~\AA\ and AIA 171~\AA\ during the strong downflow event at around 17:19~UT ($t=16$). The color of the \ha\ LOS Doppler velocity map is scaled in the range of -8\kms\ (blue) to 8\kms\ (red). Thick contours on each image represent brightenings in the IRIS~SJI~1330~\AA, and thin gray contours on the LOS Doppler velocity map indicate the outer boundary of sunspot umbra and penumbra obtained from the FISS/\ca-4~\AA\ image. \label{downflow}}
\end{figure}

In Figure~\ref{downflow}, the downflowing plasma is identified in the \ha+0.5~\AA\ image, and we find associated transient brightening in the IRIS~SJI~1330~\AA. In the \ha+0.5~\AA\ image, we can see an elongated absorption feature in the center of the image. The feature is also prominent in the \ha\ LOS velocity map with positive velocity. As seen in the \ha+0.5~\AA\ image and the LOS velocity map, the downflow plasma is coming from the outer part of the sunspot to the umbra. During the downflow event, a brightening is identified in the IRIS~SJI~1330~\AA\ marked with a white arrow. The brightening corresponds to the end of the \ha\ flow, and the size of the brightening is about 3~\arcsec. As seen in the Figure~\ref{wt}~(d), the bright feature lasted about three minutes.

On the other hand, there is no associated brightening in the EUV images. In the \textit{SDO}/AIA 171~\AA\ images in Figure~\ref{downflow} and other EUV images, we cannot find any cospatial and cotemporal brightening associated with the flow identified in the \ha+0.5~\AA\ image. This means that the temperature of the plasma downflowing along the magnetic fields is not enough to emit EUV light. As a matter of fact, we find a dark elongated feature which is located on the position of the \ha\ flow, and it also corresponds to the elongated bright features in the IRIS~SJI~1330~\AA.

\section{DISCUSSION}

For the first time, we detected impulsively generated three-minute oscillations above a sunspot in the chromospheric and TR lines. These oscillations are distinct from the three-minute umbral oscillations which exist persistently \citep[e.g.,][]{bry99}. The oscillations we detected were driven by a strong downflow event above a sunspot. The downflow event suddenly appeared in the \ha, \cb\ and \si\ lines, and then the oscillations were identified in all of the lines. The amplitudes of the Doppler velocity oscillations were about 2~\kms, and gradually decreased with time. In addition, the oscillation period was initially 2.7~minutes, and gradually increased to 3.3~minutes. Note that the Doppler velocity oscillations showed a similar behavior in all of the lines after the downflow event. That is to say, the amplitudes of the oscillations were the same values, and there were no evident phase differences among the five lines. We also found that a transient brightening in the IRIS~SJI~1330~\AA\ image corresponds to the footpoint of the downflowing material shown in the \ha+0.5~\AA\ image.

The characteristics of the downflowing material are similar to those of sunspot plumes to a certain degree. Sunspot plumes are generally located above sunspots and have downflows of 20--30~\kms\ \citep{mal99, bry01, bro05}. They flow from outside the sunspots toward the umbrae due to gas pressure difference \citep{doy03}. In addition, it has been reported that plumes are identified in EUV lines formed at TR temperatures of $5.2\leq\log{T}$(K)$\leq5.9$ \citep{bro05}. However, the downflowing material we observed is found in the \ha\ line and only the footpoint part of the material is identified in the FUV line of IRIS~SJI~1330~\AA. Since the downflowing material is identified in the \ha\ line and not in the EUV lines, it may imply that the temperature of the material is lower than that of the typical sunspot plumes.

We conjecture that the three-minute oscillations we observed represent gravity-modified acoustic waves generated by an impulsive disturbance \citep{lamb09, kal94, chae15}. The well-known work of \citet{fle91} demonstrated that three-minute oscillations arise as a natural response of the atmosphere excited by a five-minute piston motion in the lower boundary. \citet{chae15} regarded this phenomenon as a consequence of the sudden set-up of the driving motion which is a sort of impulsive disturbance from the initial equilibrium. According to \citet{chae15} an impulsive disturbance in a gravitationally stratified medium generates two packets of dispersive acoustic waves, i.e., high frequency waves and low frequency waves. Since the group speeds of the high frequency waves are higher than those of the low frequency waves, they propagate faster than the low frequency waves. For this reason, at a fixed position, the observed oscillation frequency changes from high to low values, or the oscillation period changes from short to long values. In addition, since the waves carry energy away, the amplitude of the oscillations decreases with time. These theoretical characteristics are consistent with our findings of the occurrence of the oscillations, the increase of the oscillation period and the decrease of the oscillation amplitude. Indeed, the waves generated in a gravitationally stratified medium are dispersive, and they propagate out of the source region \citep{kal94, chae15}. Thus, there should be a phase difference of the velocity oscillations when they are observed in two different atmospheric layers. Additionally, the amplitudes of the velocity oscillations may be different in the two different layers, since they are affected by physical conditions of the layer which they belong to.

One may wonder why the oscillations in our data show a similar behavior in all of the chromospheric (\ca, \ha, \mg) and the TR lines (\cb, \si). A possible explanation is that the five spectral lines were formed in the same atmospheric layer. The bright feature seen in the IRIS~SJI~1330~\AA\ corresponds to the end of the downflow material. It indicates that the footpoint of the flow rooted in the lower atmosphere (chromosphere) is heated up to middle TR temperatures ($10^{4.9}$~K) probably by the gravitational energy of the material flowing along the flux tube. \textbf{The TR temperature plasma, however, may be far from ionization equilibrium. Hence all the neutrals and singly-ionized elements will get ionized in a short time because of the enhanced temperature and density \citep{carl02}. The TR temperature plasma, however, may still contain some amount of neutrals and singly-ionized elements reproduced by the recombination of ions and electrons. The spectral lines of these species may display the same dynamical property as the other TR lines, and therefore the oscillations we identified in the five spectral lines may show high similarities.}

Our results suggest that impulsive events can drive three-minute oscillations in the chromosphere. Among them we observed a transient downflow event that occurred in the chromospheric parts of magnetic loops anchored inside the sunspot. Since the solar chromosphere displays a variety of impulsive events, we expect that all the events may significantly contribute to the generation of oscillations and waves in the solar chromosphere.

\acknowledgments
We appreciate the referee's constructive comments. This work was supported by the National Research Foundation of Korea (NRF - 2012 R1A2A1A 03670387). NST operation is partly supported by the Korea Astronomy and Space Science Institute and Seoul National University, and by stategic priority research program of CAS with Grant No. XDB09000000. Y.-H. Kim is supported by the ``Maintenance of Space Weather Research Center", a project of KASI. Y.-H. Kim and E.-K.L. are supported by the ``Planetary system research for space exploration" from KASI. M.M. is funded by the Leverhulme Trust, UK. IRIS is a NASA small explorer mission developed and operated by LMSAL with mission operations executed at the NASA Ames Research center and major contributions to downlink communications funded by the Norwegian Space Center (NSC, Norway) through an ESA PRODEX contract.

\clearpage

\clearpage

\clearpage

\clearpage

\clearpage


\begin{thebibliography}{}

\bibitem[Aschwanden et al.(1999)]{asc99} Aschwanden, M.~J.,
Fletcher, L., Schrijver, C.~J., \& Alexander, D.\ 1999, \apj, 520, 880
\bibitem[Beckers \& Schultz(1972)]{bec72} Beckers, J.~M., \& Schultz, R.~B.\ 1972, \solphys, 27, 61
\bibitem[Beckers \& Tallant(1969)]{bec69} Beckers, J.~M., \& Tallant, P.~E.\ 1969, \solphys, 7, 351
\bibitem[Bogdan \& Judge(2006)]{bog06} Bogdan, T.~J., \& Judge, P.~G.\ 2006, Philosophical Transactions of the Royal Society of London Series A, 364, 313
\bibitem[Brosius(2005)]{bro05} Brosius, J.~W.\ 2005, \apj, 622, 1216
\bibitem[Brynildsen et al.(2004)]{bry04} Brynildsen, N., Maltby, P., Foley, C.~R., Fredvik, T., \& Kjeldseth-Moe, O.\ 2004, \solphys, 221, 237
\bibitem[Brynildsen et al.(2001)]{bry01} Brynildsen, N., Maltby, P., Fredvik, T., Kjeldseth-Moe, O., \& Wilhelm, K.\ 2001, \solphys, 198, 89
\bibitem[Brynildsen et al.(1999)]{bry99} Brynildsen, N.,
Leifsen, T., Kjeldseth-Moe, O., Maltby, P., \& Wilhelm, K.\ 1999, \apjl, 511, L121
\bibitem[Carlsson \& Stein(2002)]{carl02} Carlsson, M., \& Stein, R.~F.\ 2002, \apj, 572, 626
\bibitem[Chae \& Goode(2015)]{chae15} Chae, J., \& Goode, P.~R.\ 2015, \apj, 808, 118
\bibitem[Chae et al.(2013)]{chae13} Chae, J., Park, H.-M., Ahn, K., et al.\ 2013, \solphys, 288, 1
\bibitem[De Pontieu et al.(2014)]{dep14} De Pontieu, B.,
Title, A.~M., Lemen, J.~R., et al.\ 2014, \solphys, 289, 2733
\bibitem[Deubner et al.(1996)]{deub96} Deubner, F.-L., Waldschik, T., \& Steffens, S.\ 1996, \aap, 307, 936
\bibitem[Doyle \& Madjarska(2003)]{doy03} Doyle, J.~G., \& Madjarska, M.~S.\ 2003, \aap, 407, L29
\bibitem[Fleck \& Schmitz(1991)]{fle91} Fleck, B., \& Schmitz, F.\ 1991, \aap, 250, 235
\bibitem[Giovanelli(1972)]{gio72} Giovanelli, R.~G.\ 1972, \solphys, 27, 71
\bibitem[Kalkofen et al.(1994)]{kal94} Kalkofen, W., Rossi, P., Bodo, G., \& Massaglia, S.\ 1994, \aap, 284, 976
\bibitem[Khomenko \& Collados(2015)]{kho15} Khomenko, E., \& Collados, M.\ 2015, Living Reviews in Solar Physics, 12, 6
\bibitem[Kosovichev \& Zharkova(1998)]{kos98} Kosovichev, A.~G., \& Zharkova, V.~V.\ 1998, \nat, 393, 317
\bibitem[Kosovichev(2006)]{kos06} Kosovichev, A.~G.\ 2006, \solphys, 238, 1
\bibitem[Lamb(1909)]{lamb09} Lamb, H.\ 1909, Proc. London Math. Soc., 7, 122
\bibitem[Lemen et al.(2012)]{lemen12} Lemen, J.~R., Title,
A.~M., Akin, D.~J., et al.\ 2012, \solphys, 275, 17
\bibitem[Lites \& Thomas(1985)]{lites85} Lites, B.~W., \& Thomas, J.~H.\ 1985, \apj, 294, 682
\bibitem[Maltby et al.(1999)]{mal99} Maltby, P., Brynildsen,
N., Fredvik, T., Kjeldseth-Moe, O., \& Wilhelm, K.\ 1999, \solphys, 190, 437
\bibitem[Nakariakov et al.(1999)]{nak99} Nakariakov, V.~M.,
Ofman, L., Deluca, E.~E., Roberts, B., \& Davila, J.~M.\ 1999, Science, 285, 862
\bibitem[O'Shea et al.(2002)]{osh02} O'Shea, E., Muglach, K., \& Fleck, B.\ 2002, \aap, 387, 642
\bibitem[Schrijver et al.(2002)]{sch02} Schrijver, C.~J.,
Aschwanden, M.~J., \& Title, A.~M.\ 2002, \solphys, 206, 69
\bibitem[Staude(1999)]{sta99} Staude, J.\ 1999, in ASP Conf. Ser. 184, Third Advances in Solar Physics Euroconference: Magnetic Fields and Oscillations, ed. B. Schmieder, A. Hofmann, \& J. Staude (San Francisco: ASP), 113
\bibitem[Thomas et al.(1987)]{tho87} Thomas, J.~H., Lites,
B.~W., Gurman, J.~B., \& Ladd, E.~F.\ 1987, \apj, 312, 457
\bibitem[Tian et al.(2014)]{tian14} Tian, H., DeLuca, E.,
Reeves, K.~K., et al.\ 2014, \apj, 786, 137
\bibitem[Torrence \& Compo(1998)]{tor98} Torrence, C., \& Compo, G.~P.\ 1998, Bulletin of the American Meteorological Society, 79, 61
\bibitem[Zimovets \& Nakariakov(2015)]{zim15} Zimovets, I.~V., \& Nakariakov, V.~M.\ 2015, \aap, 577, A4


\end{thebibliography}
\end{document}